\documentclass[aps,prd,twocolumn,floatfix]{revtex4} 
\usepackage{epsfig} 
 
\def\gsim{ \lower .75ex \hbox{$\sim$} \llap{\raise .27ex \hbox{$>$}} }  
\def\lsim{ \lower .75ex \hbox{$\sim$} \llap{\raise .27ex \hbox{$<$}} } 
\def\be{\begin{equation}} 
\def\ee{\end{equation}}

\def\Vbar{V}

\def\mphi{m_{\Phi}} 
\def\mpl{M_{Pl}}

\begin{document} 
 
\title{Tuning Locked Inflation: Supergravity versus Phenomenology} 
 
\author{Richard Easther$^{1,2}$} 
\author{Justin Khoury$^2$} 
\author{Koenraad Schalm$^2$}
 
\affiliation{~}
\affiliation{$^1$Department of Physics, Yale University, New Haven  CT 06520, USA}
\affiliation{$^2$ISCAP, Pupin Hall, Columbia University, New York~NY~10027, USA} 
 
\begin{abstract} 
We analyze the cosmological consequences of {\it locked inflation\/},
a model recently proposed by Dvali and Kachru that can produce
significant amounts of inflation without requiring slow-roll. 
We pay particular attention to the end of inflation in this model,
showing that a 
secondary phase of {\it
  saddle inflation\/} can follow the locked
inflationary era.  However, this subsequent period of inflation results in a
strongly scale dependent spectrum that can lead to massive black hole
formation in the primordial universe. Avoiding this disastrous outcome 
puts strong constraints on the parameter space open to models of locked inflation. 
\end{abstract} 
 
\maketitle 

\section{Introduction}

The inflationary paradigm~\cite{inf} provides a compelling account of
early universe cosmology. The universe emerges from the inflationary
phase with large-scale homogeneity and endowed with a nearly scale
invariant spectrum of density fluctuations, consistent with current
observations. Despite these impressive phenomenological achievements,
designing successful models of inflation within supergravity and
string theory has proven to be a frustratingly difficult
task~\cite{lyth}. In spontaneously broken supergravities
lifted flat directions are natural inflaton candidates.
However, the various moduli 
have
(stable, protected) masses $m$ of order $H$, the Hubble constant
during inflation~\cite{note,nima}. 
This perversely spoils slow-roll inflation
since the slow-roll condition, $\eta \sim m^2/H^2\ll 1$, is then
violated.  In other words, the generic outcome in supergravity
theories is $\eta\;\gsim\; {\cal O}(1)$. This so-called
$\eta$-problem is encountered, for instance, in attempts to embed
inflation in the stringy landscape~\cite{mald}. 
Recent developments, however, indicate
that a simple change to the K\"ahler potential might alleviate this problem~\cite{newstuff}.

In a recent paper, Dvali and Kachru \cite{Dvali:2003vv} (henceforth,
DK) introduced {\em locked inflation\/} as a possible way out of this
dilemma. 
Its distinguishing feature is that it does away with the
slow-roll constraints. Instead locked inflation relies on the rapid oscillations of
one scalar field which lock a second field at the top of a saddle
point. The potential energy at the saddle then drives inflation. 
At the very least, this model is an intriguing alternative to slow-roll
inflation and, if consistent, overcomes the hurdles faced by slow-roll inflation in supergravity and string
theories. Locked inflation needs no intrinsically small parameters,
although it does exploit -- like most two-field models -- the ratio of
the two widely separated scales. The existence of widely
separated scales is not a new tuning, however, as this hierarchy must be
explained even in the absence of inflation.  

In this paper, 
we examine the termination of locked inflation and the
subsequent evolution of the universe. It is perhaps natural to assume 
that inflation ends as soon as the field-point is no longer trapped at the 
saddle point.  However, 
for the parameter values natural in broken supergravities,
we find that this is not necessarily the case. Instead, the universe can undergo a second
period of 
inflation as the field point moves orthogonally to
the direction about which it was previously oscillating. We dub this
phase {\it saddle inflation\/}.  

Saddle inflation has potentially disastrous observational consequences
for the DK scenario.  As we will show, modes that leave the horizon at the onset of saddle inflation
typically have an amplitude of order unity and thus give rise to a
phenomenologically dangerous number of massive black holes when they re-enter the horizon during the subsequent radiation or matter-dominated eras.  
The formation of these black holes must thus be
avoided at all costs.
There are two ways to do so. 
One is to demand that there be no saddle inflation at all.
This requires that the (tachyonic) mass of the
saddle field be much larger than $H$, or $\eta\gg 1$. 
The other way out of the black hole problem is to make the secondary phase of 
saddle inflation last long
enough to move the dangerous range of scales outside the present
cosmological horizon. This renders the prior period of locked inflation
unobservable, although the latter remains useful as a mechanism for resolving
the initial conditions problem related to the onset of inflation. 
Long saddle inflation naturally occurs for $\eta \ll 1$.  

The generation of large perturbations at a saddle point of the potential
has been discussed for general two-field models
by Garcia-Bellido, Linde and Wands \cite{Wands:1996},
and in the specific context of supernatural inflation 
by Randall, Soljacic and Guth
\cite{Randall:1995}. 
 Adopting the analyses of these earlier models to the specific case of
locked inflation, we can readily deduce
the phenomenological consequences associated with the end of locked inflation.

We conclude that a viable model of locked
inflation requires either $\eta\;\gsim\; 30-1500$ (no saddle inflation), depending on the
reheating temperature, or $\eta\; \lsim \; 0.01-2.5$ (long saddle inflation), 
similarly depending on the reheating temperature. 
Thus, at a naive level, it appears that both models necessitate a
similar degree of tuning. 
 
It is noteworthy that long saddle inflation with $\eta\sim {\cal O}(1)$ is allowed for reheating temperature in the range
$1-10^9$~GeV. Locked inflation only sets up the desired initial
condition for saddle inflation in this case. Nevertheless, this is
a successful 
inflationary model with supergravity-inspired potential. As
such, it is a clear candidate for embedding
inflation in string theory and supergravity. 
To reproduce the observed scale invariant spectrum of density perturbations,
however, it does require that the fluctuations arise from an alternative mechanism.

\section{Locked inflation: a review} 
 
Consider the simple, two-field potential 
\be 
{\cal V}(\Phi,\phi) = \mphi^2 \Phi^2 + \lambda \Phi^2 \phi^2 + 
M^4\left(1 - \frac{\eta}{4}\frac{\phi^2}{M_{Pl}^2}\right)^2\,, 
\ee 
where $m_{\Phi}^2\sim M^4/M_{Pl}^2$, $\lambda$ is a dimensionless
parameter of order unity, $M$ is of order of the supersymmetry
breaking scale, and $M_{Pl} = (8\pi G)^{-1/2}$ is the reduced Planck
mass. The meaning of the dimensionless parameter $\eta$ will soon
become clear~\cite{compare}. DK show that this potential produces
locked inflation when the field-point oscillates rapidly in $\Phi$
direction, while $\phi \sim 0$. Eventually these oscillations are
sufficiently damped to allow the field point to roll off the saddle
point in the $\phi$ direction, thereby ending locked inflation.   
 
As the universe inflates, the Hubble parameter $H$ is nearly constant
and given by $3H^2M_{Pl}^2\approx M^4$.
The number of e-folds of inflation generated while $\Phi$ oscillates
equals
\begin{equation} 
N_{locked}\approx \frac{4}{3}\ln\left(\frac{M_{Pl}}{M}\right)\,. 
\label{Nlocked} 
\end{equation} 
For $M = 1$ TeV, this gives $N_{locked}\approx 50$, and therefore one
stage of locked inflation is sufficient in this case
to account for the homogeneity and isotropy of our observable
universe. Larger values of the small energy
scale $M$, however, require multiple stages, each
contributing $N_{locked}$ e-folds of inflation. 
This ``cascade'' scenario is not unreasonable in string theory, as DK argue, since we
expect that the field point will go through many saddle points in the
stringy landscape before reaching the true vacuum of the theory.  
 
Our numerical calculations confirm the estimate for $N_{locked}$ given
above. However, there is an additional constraint on $m_\Phi$. If it
is too small ($< 3H/2\sqrt{2}$),  $\Phi$ will be overdamped, and no
oscillations will occur.  On the other hand, if $m_\Phi$ is too large
($> 10H/\sqrt{2}$), one can produce $\Phi$ particles via parametric resonance,
and the kinetic energy will be rapidly drained from the $\Phi$ field,
undermining locked inflation~\cite{param}. While this restricts the
parameter range open to successful models of locked inflation, the
natural expectation from supergravity that $m_\Phi \sim H$ is
compatible with these constraints. 

\section{Inflation After Locked Inflation} \label{phiinf} 
 
In the remainder of this paper, we focus on the field evolution
immediately after locked inflation. In particular, we argue that it is
possible for the universe to undergo further inflation for a wide
range of parameter values. This effect is well known in two-field
models \cite{Wands:1996} and is particularly important in supernatural
inflation~\cite{Randall:1995} where it produces a large ``spike'' in
the density perturbation spectrum at small wavelengths.  
 
Once a given stage of locked inflation ends, the field point rolls off
in the $\phi$-direction. By this time the oscillations in $\Phi$ can
safely be ignored and, since $\Phi\approx 0$, the problem reduces
effectively to a single-field model with potential  
\begin{equation} 
\Vbar(\phi) = M^4\left(1-\frac{\eta}{4}\frac{\phi^2}{M_{Pl}^2}\right)^2\,. 
\label{Vbar} 
\end{equation} 

We derive conditions under which a subsequent inflationary phase will
occur as $\phi$ rolls off, a phase which we henceforth refer to as
{\it saddle inflation}. In general this phase will not satisfy the
usual slow roll conditions of inflation, however. Indeed, near the top
of the hill where $\phi\ll M_{Pl}$, the slow-roll parameter $\eta_{s}$
is 
given by  
\begin{equation} 
|\eta_{s}|\equiv \mpl^2\left\vert\frac{V_{,\phi\phi}}{V}\right\vert =
 \eta +{\cal O}\left(\frac{\phi}{M_{Pl}}\right)\,, 
\label{eta} 
\end{equation} 
which is greater than unity for $\eta\;\gsim\; 1$.  The virtue of writing the potential in Eq.~(\ref{Vbar}) in terms of $\eta$ is now clear, as the latter reduces to $|\eta_{s}|$ for small $\phi$. 
 
While the usual slow-roll expressions are not generally applicable
here, we can nevertheless make use of a quadratic approximation. 
In natural inflation~\cite{Freese:1990rb,Adams:1992bn}
this is known as the {\it small angle\/} approximation. Its range
of applicability is much greater than that of the slow-roll
approximation, and  it contains slow-roll as a limit. For small
$\phi$, $\Vbar$ reduces to  
\begin{equation} 
\Vbar(\phi)\approx M^4 - \frac{1}{2}\frac{\eta M^4}{M_{Pl}^2}\phi^2\,. 
\label{quadpot}
\end{equation} 
The dynamics of $\phi$ are thus governed by the equation of motion 
(recall that we approximate $H$ to be constant)  
\begin{equation} 
\frac{d^2\phi}{dN^2} + 3\frac{d\phi}{dN} \approx  3\eta\phi\,, 
\label{eomphi}
\end{equation} 
where $N\equiv \ln a$ is the number of e-folds. The initial conditions
for $\phi$ can be estimated as follows. At the end of the locked
inflation, the effective mass in the $\phi$ direction is effectively
zero, but the expected quantum fluctuation in $\phi$ is on the order
of $H$. 
Thus, semiclassically we can assume $\phi_{init}\sim H$~\cite{Wands:1996} and
that the initial velocity is zero. With these initial conditions, the
solution to Eq.~(\ref{eomphi}) is 
\begin{eqnarray}
\phi(N) &=& \frac{\phi_{init}}{2\delta} \left\{ 
(\delta +1) \exp{\left[\frac{3}{2} (\delta -1) N \right] } + \right. 
\nonumber \\
&& \left. (\delta -1) \exp\left[-\frac{3}{2}(\delta +1) N \right] \right\}\,,
\label{phifull}  
\end{eqnarray}
where $\delta \equiv \sqrt{1+ 4\eta/3}$. 
Let us assume that we will be able to show that 
this solution can correspond to an inflationary phase, which by
definition lasts longer than one e-fold.
Since $\delta > 1$,
after one e-fold we can ignore the second term in Eq.~(\ref{phifull})
and write 
%
\be
\phi \approx 
\frac{\phi_{init}}{2\delta} (\delta+1) 
\exp(f(\eta)N)\,,
\label{phiexp}
\ee
where
\begin{equation} 
f(\eta)\equiv \frac{3}{2}\left\{\sqrt{1+\frac{4}{3}\eta}-1\right\}\,. 
\label{f} 
\end{equation} 
The function $f$ will play an important role in the results to be
derived below.  

To show that this solution indeed yields 
an inflationary phase, consider  
\begin{equation} 
\epsilon = \frac{3}{2}(1+w)\,, 
\end{equation} 
where $w$ is the equation of state for $\phi$, given by $w \equiv
(\dot{\phi}^2-2V)/(\dot{\phi}^2 + 2V)$. In the slow-roll
approximation, $\epsilon$ reduces to the usual slow-roll parameter
$\epsilon_{s}\equiv M_{Pl}^2V_{,\phi}^2/2V^2$. Inflation takes place
if $w\approx -1$, or, equivalently, $\epsilon\ll 1$. Using
Eq.~(\ref{phiexp}), $\epsilon$ is seen to equal
\begin{equation}
\epsilon \approx 
\frac{1}{2}
f^2(\eta)\frac{\phi^2}{M_{Pl}^2}\,. 
\label{bareps} 
\end{equation} 

\begin{figure}[tbp]
\centering
\epsfig{file=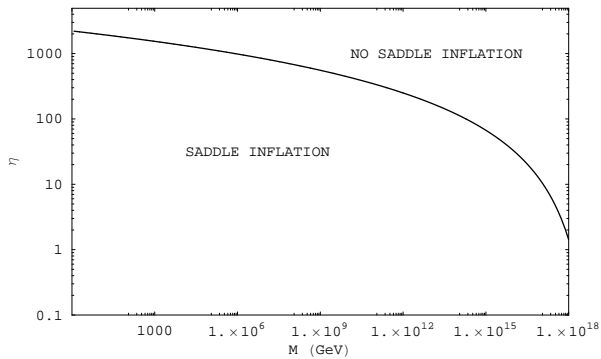,width = 8cm}
\caption[]{The curve delimits those values of $\eta$ and $M$~(GeV) that lead to saddle inflation from those that do not.}
\label{cons}
\end{figure}

Saddle inflation terminates when $\epsilon\sim {\cal O}(1)$ or when the quadratic
approximation breaks down, whichever happens first. For $\eta\;\gsim\;1$, the relevant
range for the discussion below, inflation ends when $\epsilon\sim {\cal O}(1)$.
From Eq.~(\ref{bareps}), this occurs when the field reaches
$\phi_{end}\approx \sqrt{2}f^{-1}(\eta)M_{Pl}$.
Substituting in Eq.~(\ref{phiexp}), we find that the total number of e-folds,
$N_{saddle}$, generated during saddle inflation is given by   
%
\begin{eqnarray}
\nonumber
N_{saddle} &=& \frac{1}{f(\eta)}\ln\left(\frac{\sqrt{2}(2f(\eta)+3)}{f(\eta)(f(\eta)+3)} \frac{M_{Pl}}{\phi_{init}}\right)\\
&\approx &
\frac{1}{f(\eta)}\ln\left(\frac{2\sqrt{6}}{f(\eta)}\frac{M_{Pl}^2}{M^2}
\right)\,.   
\label{Nsaddle2} 
\end{eqnarray} 
 
The discussion thus far only applies if $N_{saddle}\;\gsim\; 1$;
otherwise there is no inflation. Conversely, there will be no
saddle inflation if $N_{saddle}\;\lsim\; 1$. Using
Eqs.~(\ref{Nsaddle2}) and~(\ref{f}) we can derive a necessary and
sufficient condition for the existence of a period of saddle
inflation by working out the minimum value of $\eta$ for which
$N_{saddle}<1$.  This bound is plotted in Fig.~\ref{cons}.
As illustrative examples, the bound on $\eta$ for two physically
interesting values of $M$ is: 
\begin{eqnarray}
\nonumber 
& & \eta_{min}\;\approx\; 1500 \qquad {\rm for}\;\;\; M\sim 1\; {\rm TeV}\\ 
& & \eta_{min}\;\approx\; 30 \qquad \;\;\;\;{\rm for}\;\;\; M\sim 10^{16}\; {\rm GeV}\,. 
\label{etalocked} 
\end{eqnarray} 
 
To summarize, we have shown that some inflation will follow the period of locked inflation if $\eta \;\lsim \; 10-10^{3}$, depending on the choice of 
reheating temperature $M$.

\section{Perturbation spectrum and black hole formation} \label{bh} 
 
We next compute the spectrum of density fluctuations generated during
saddle inflation. In Newtonian gauge, the linearized equation for the
$k$-mode $u_k$ of the gauge-invariant perturbation variable $u$,
related to the Newtonian potential by $\Phi_{Newton}=uH(d\phi/dN)$,
is~\cite{Gratton:2003pe}  
\begin{equation} 
\frac{d^2u_k}{dN^2} + \frac{du_k}{dN} + \left(\frac{k^2}{a^2H^2} - \beta(N) \right) u_k = 0\,; 
\label{u2} 
\end{equation} 
\begin{eqnarray} 
\nonumber
\beta(N) &=& 
\frac{1}{2}\left\{\frac{d\ln \epsilon}{dN} + 
\frac{1}{2}\left(\frac{d\ln \epsilon}{dN}\right)^2 - \frac{d^2\ln 
\epsilon}{dN^2}\right\} + {\cal O}\left(\epsilon\right)
\,,  
\label{m} 
\end{eqnarray} 
where we recall that $a(N)\equiv \exp(N)$. The only approximation made in deriving Eq.~(\ref{u2}) is that the universe is inflating, that is, $\epsilon\ll 1$. No assumption was made about the time-dependence of $\epsilon$, however. 
 
Substituting Eq.~(\ref{bareps}) and using Eq.~(\ref{phiexp}), we find 
\begin{equation} 
\beta \approx f(\eta)(1+f(\eta))\,. 
\end{equation} 
In particular, since $\beta$ is nearly constant, Eq.~(\ref{u2}) becomes analytically solvable. Choosing the usual Bunch-Davies vacuum in the short-wavelength limit, we get~\cite{Gratton:2003pe} 
\begin{eqnarray} 
\nonumber 
&& k^{3/2}\,u_k = \frac{\sqrt{\pi}}{4}\sqrt{\frac{k}{aH}} 
  H_{p}^{(1)}\left(\frac{k}{aH}\right) 
\\
& & p(\eta) \equiv \sqrt{f(\eta)(1+f(\eta)) + 1/4}\,, 
\label{soln} 
\end{eqnarray} 
where $H_p^{(1)}$ is the Hankel function of the first kind of order $p$.
In the long-wavelength limit, $k^2/a^2H^2\ll \beta$, this gives 
\begin{equation} 
k^{3/2}|u_k| \approx \frac{\sqrt{\pi}}{2^{2-p}\sin(\pi p)\Gamma(1-p)} \left(\frac{k}{aH}\right)^{-p+1/2}\,, 
\label{long} 
\end{equation} 
from which we can read off the spectral index of the fluctuations: 
\begin{equation} 
n_s - 1 \approx -2p(\eta)+1 = -\sqrt{1+4f(\eta)(1+f(\eta))} + 1\,. 
\label{tilt} 
\end{equation} 
As a check, recall that in the slow-roll approximation we have
$f(\eta)\approx \eta \ll 1$, and thus $n_s-1 \approx -2\eta$. We 
see from Eq.~(\ref{eta}) that this agrees with the usual slow-roll
expression for the spectral index $n_s$. 

The amplitude of the density perturbations 
is naturally expressed in terms of $\zeta_k$, the curvature perturbation on comoving
 hypersurfaces~\cite{bardeen}. This gauge-invariant variable is related to $u_k$ by 
\begin{equation} 
\zeta_k = \frac{H}{\epsilon a}\frac{d}{dN}\left(a\frac{d\phi}{dN}u_k\right)\,. 
\label{zeta} 
\end{equation} 
In single-field inflation $\zeta_k$ is nearly constant outside the
horizon. Single-field inflation is a good approximation
during saddle inflation since, as mentioned earlier, the second field,
$\Phi$, is essentially inert. Thus it suffices to evaluate $\zeta_k$
at horizon-crossing, that is, at $k = aH$. Substituting
Eq.~(\ref{long}), and using Eqs.~(\ref{phiexp}) and~(\ref{bareps}), we
obtain  
\begin{equation} 
k^{3/2}\zeta_k = \frac{\sqrt{\pi}[f(\eta) + p(\eta) + 1/2]}{2^{2-p}\sin(\pi p)\Gamma(1-p)}
\left. \frac{\sqrt{2}HM_{Pl}} {\sqrt{\epsilon}}\right|_ {k=aH}\,, 
\label{final} 
\end{equation} 
where the subscript ``$k=aH$'' denotes horizon-crossing.  
 
Let us evaluate this quantity for the first mode to leave the horizon
during saddle inflation. This mode freezes out when
$\phi=\phi_{init}\sim H$, and, therefore, from Eq.~(\ref{bareps}),
$\sqrt{\epsilon}_{k=aH} \approx f(\eta) H/\sqrt{2}M_{Pl}$.
Substituting in Eq.~(\ref{final}), we find that it has an amplitude 
\begin{eqnarray} 
\frac{(k^{3/2}\zeta_k)_{init}}{M_{Pl}^2} &=& \frac{\sqrt{\pi}[f(\eta) + p(\eta) + 1/2]}{2^{2-p(\eta)}\sin(\pi p(\eta))\Gamma(1-p(\eta))}
\frac{2H}{\phi_{init}f(\eta)}
\label{longwave} 
\nonumber
\\
&\approx &\frac{\sqrt{\pi}[f(\eta)+p(\eta)+1/2]}{2^{2-p(\eta)}
\sin(\pi p(\eta))\Gamma(1-p(\eta))} \frac{2}{f(\eta)} \, . 
\end{eqnarray} 
Since $p$ and $f$ are both functions of $\eta$, this entire expression
depends solely on $\eta$. Moreover, it is easily seen that it is
numerically always greater than unity. 
 
Therefore, over the range of scales corresponding to the first few
modes to leave the horizon during saddle inflation, the amplitude of
density fluctuations is of order unity. When these modes re-enter the
horizon after reheating, there is a probability of roughly a half that
the overdensity will collapse and form a black hole~\cite{carr}. The
issue of black hole formation is not unique to locked inflation, and
has been been studied 
in the context of supernatural
inflation in particular~\cite{Wands:1996}. 
In the case of supernatural inflation, the problem of black hole formation
can be avoided in two ways: either $\eta$ must be tuned to sufficiently large values
to prevent saddle inflation from occurring; or the scale of
inflation must be sufficiently high so that the black holes evaporate
well before nucleosynthesis. 
In the context of locked inflation, as we will
now argue, this latter option fails.
 
\section{Consequences for locked inflation} 
 
Consider first the model 
where
a single stage of locked inflation is sufficient to solve the horizon
and flatness problems, {i.e.}
we choose $M\sim 1$ TeV. 
Suppose that this stage of
locked inflation is
followed by a period of saddle inflation, which, from
Eq.~(\ref{etalocked}), requires $\eta\;\lsim\; 1600$. As mentioned
above, when the mode corresponding to $\phi_{init}$ re-enters the
horizon, a black hole 
is likely to form. It suffices
to focus on the case where only one e-fold or so is generated during
saddle inflation since a longer inflationary phase will only make the
black hole larger. Then, the Schwarzschild radius of the black hole
formed, $R_S\sim\delta\rho/H^{3}M_{Pl}^2$, is of the order of the Hubble radius at reheating, $H^{-1} \sim M_{Pl}/M^2$. 
That is,  the black hole has mass
\begin{equation} 
M_{BH} \sim \frac{M_{Pl}^3}{M^2}\,. 
\label{mbh1} 
\end{equation} 
For $M\sim 1$ TeV, this gives $M_{BH} \sim 10^{26}$ g, or roughly the
mass of the Earth! These massive black holes have a lifetime longer
than the present age of the universe and thus dominate the 
evolution of the 
universe
well before nucleosynthesis.

Equation~(\ref{mbh1}) suggests that this problem could be avoided
by making the small scale
$M$ sufficiently large. Indeed, it seems 
this would make $M_{BH}$ 
sufficiently small such that the black holes
would evaporate well before nucleosynthesis. This is indeed the avenue
taken in supernatural inflation where one imposes $M\;\gsim\; 10^{11}$
GeV~\cite{Wands:1996}.   
 
However, locked inflation is different. 
This is because larger values of
$M$ require multiple
 stages of locked inflation, as in the cascade model
proposed by DK. Suppose each stage generates $\Delta N_{locked}$
e-folds of inflation, with $\Delta N_{locked}$ given by
Eq.~(\ref{Nlocked}).  Moreover, suppose that a substantial fraction
of all stages are followed by a short period of saddle inflation. As argued
in Sec.~\ref{bh}, to each phase of saddle inflation there corresponds
a range of scales for which black holes will form when the
corresponding modes re-enter the horizon. 
The typical wavelength of the largest black hole thus formed will be
of order  
\begin{equation} 
M_{BH} \sim \frac{M_{Pl}^2}{H_0}\left(\frac{M}{M_{Pl}}\right)^{8/3}\,, 
\end{equation} 
where $H_0$ is today's Hubble constant. 
For $M =10^{11}$~GeV, say, this gives
$M_{BH} \approx 10^{35}$~g. 
Therefore, in locked inflation with short stages of 
saddle inflation, the generic mass of black holes
produced {\em increases} with $M$, making the problem worse. 

\section{How to avoid black hole formation}

\begin{figure}[tbp]
\centering
\epsfig{file=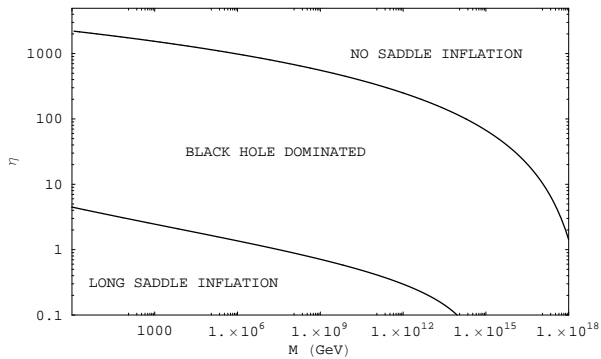,width = 8cm}
\caption[]{Phenomenologically allowed choices for $\eta$ and $M$~(GeV). 
The region ``Long Saddle Inflation'' corresponds to the values of 
$\eta$ and $M$ for which saddle inflation lasts long enough to push the 
dangerous range of modes outside the present horizon. Unacceptable black hole production at horizon reentry excludes the middle region. 
\label{cons2}}  
\end{figure}

As mentioned earlier,
one way out of the black hole problem is simply to avoid 
any saddle inflation. This is the case if $(\eta,M)$ lie in the ``No Saddle Inflation'' region of Fig.~\ref{cons}.

An alternative approach is to impose that saddle inflation lasts long enough to push the dangerous range of modes well beyond the scale of our observable universe. In other words, modes with amplitude of order unity are still generated, but their wavelength is sufficiently long that they have not re-entered the horizon by the present time. In this case, however, all the modes within our observable universe are generated during saddle inflation, leaving locked inflation with no observational consequences.

In order to be phenomenologically viable, the long period of saddle inflation must satisfy two constraints. First, as mentioned above, it must last sufficiently long to push the dangerous modes outside our observable universe. Second, the spectrum of perturbations within our horizon must be consistent with observations. 
This latter constraint normally forces one to the slow-roll regime ($\eta\ll 1$). 
However, quantum fluctuations of the ``inflaton'' $\phi$ 
need not necessarily be responsible for the spectrum of density perturbations. In particular, let us suppose that that they are produced via the alternative mechanism of Dvali, Gruzinov and Zaldarriaga~\cite{dgz} (henceforth, DGZ).
In this case, the spectrum is generated 
by the fluctuations of some other field $\chi$. This field is assumed to be nearly massless during inflation, and, moreover, its vacuum expectation value determines the coupling constants between the inflaton and the various Standard Model fields. 

Allowing for alternative sources of density perturbations, such as the DGZ mechanism, greatly expands the range of allowed models. For instance, the spectral index for the fluctuations generated a la DGZ is~\cite{dgz}
\begin{equation}
(n_s - 1)_\chi = -2\epsilon\,.
\label{nschi}
\end{equation}
In contrast with the familiar slow-roll result of $(n_s -1)_\phi = -6\epsilon + 2\eta$, Eq.~(\ref{nschi}) is independent of $\eta$. Since $\epsilon$ is small during saddle inflation, it is therefore possible to satisfy the observational constraint $|n_s-1|\;\lsim\; 0.1$ from WMAP~\cite{wmap} and observations of large-scale structure without requiring $\eta\ll 1$. Similarly, the expression for the amplitude of the perturbations is different than in slow-roll inflation~\cite{dgz}. For our purposes, we shall assume that it is possible for $\chi$ to lead to density perturbations with amplitude of $10^{-5}$, consistent with COBE data.

Nevertheless, fluctuations in the ``inflaton'' do contribute to the density perturbation spectrum, on top of the $\chi$ contribution. To be consistent with COBE data, these $\phi$-generated fluctuations must be less than $10^{-5}$ in amplitude on scales comparable to the size of the universe today. That is, from Eqs.~(\ref{bareps}) and~(\ref{final}), we must impose
\begin{equation}
\frac{M^2} {f(\eta)\phi_0M_{Pl}} \;\lsim \; 10^{-5}\,,
\label{conddgz}
\end{equation}
where we have neglected a factor of order unity and used the relation $H\sim M^2/M_{Pl}$.
Here $\phi_0$ is the field value of when the mode corresponding to our universe today left the horizon during saddle inflation.

Let $N_0$ denote the difference in the number of e-foldings between the mode which corresponds to our present universe and the last mode to exit the horizon during saddle inflation. This is approximately given by the logarithm of the ratio of temperatures. Since the temperature is currently $T_0=2.723$ K, while at reheating it was of order $M$, we get
\begin{equation}
N_0\approx\ln\left(\frac{M}{T_0}\right)\,.
\label{N0a}
\end{equation}
For $\eta\;\lsim\; 1$, the relevant range for this discussion, saddle inflation ends when the quadratic approximation breaks down. From Eq.~(\ref{quadpot}), this occurs when the field reaches $\phi_{end}\approx M_{Pl}/\sqrt{\eta}$. Substituting this in Eq.~(\ref{phiexp}), we obtain a second independent expression for $N_0$:
\begin{equation}
N_0\approx\frac{1}{f(\eta)}\ln\left(\frac{M_{Pl}}{\sqrt{\eta}\phi_0}\right)\,.
\label{N0b}
\end{equation}

The above expressions for $N_0$ together imply that
\begin{equation}
\frac{1}{\phi_0} \approx \frac{\sqrt{\eta}}{M_{Pl}}\left(\frac{M}{T_0}\right)^{f(\eta)}\,,
\end{equation}
which, when substituted in Eq.~(\ref{conddgz}), gives
\begin{equation}
\frac{\sqrt{\eta}}{f(\eta)}\left(\frac{M}{M_{Pl}}\right)^2\left(\frac{M}{T_0}\right)^{f(\eta)} < 10^{-5}\,.
\end{equation}
This condition on $\eta$ and $M$ ensures that the perturbations in $\phi$ are sufficiently small to be consistent with COBE data. The values of $\eta$ and $M$ which obey this condition lie in the ``Long Saddle Inflation'' region in Fig.~\ref{cons2}. 

A moment's thought reveals that this condition also ensures that saddle inflation lasts long enough to push the dangerous modes beyond our present horizon. Indeed, recall from Eq.~(\ref{tilt}) that the spectral index for the $\phi$-perturbations is given by $(n_s-1)_\phi\approx -2\eta$, corresponding to a red spectrum. That is, larger wavelength modes have larger amplitude. Moreover, recall that the dangerous modes that eventually lead to black hole formation have amplitude of order unity. Since Eq.~(\ref{conddgz}) constrains the amplitude on today's scales to be much less than unity, however, it follows that these modes must lie well beyond our horizon.

\section{Discussion} 
 
We have shown that, in order for locked inflation to be
phenomenologically viable, it must either:
 i) end without any subsequent saddle inflation;
or ii) be followed by a long phase of saddle inflation. This constrains the $(\eta,M)$
parameter space as illustrated in Fig.~\ref{cons2}. In the former case, no black holes are formed. 
In the latter case,  saddle inflation lasts sufficiently long to push the dangerous modes outside
our observable universe. However, locked inflation
would only serve the purpose of setting up the initial conditions for saddle inflation in this case, but would
have no directly testable observational consequences
in itself.

In case i), shown as ``No Saddle Inflation'' in the Figure, $\eta$ and $M$ are required to satisfy
(see Eq.~(\ref{Nsaddle2}))
%
\begin{eqnarray}
\frac{f^2(\eta)(f(\eta)+3)^2}{6(2f(\eta)+3)^2} \exp[2f(\eta)] & \gsim & \frac{M_{Pl}^4}{M^4} \nonumber \\
\stackrel{\eta \gg 1}{\Longrightarrow}~~~~~~~~~
\frac{\eta}{8} \exp[2\sqrt{3\eta}] & \gsim
&\frac{M_{Pl}^4}{M^4}\,,
\label{eq:1}
\end{eqnarray}
which implies $\eta \; \gsim \; 30-1500$ for $M \approx 10^{16}\;{\rm GeV}-1$~TeV. 

In case ii), shown as ``Long Saddle Inflation'' in the Figure,  
the bound on $\eta$ and $M$ is (see Eq.~(\ref{conddgz}))
%
\begin{eqnarray}
  \label{eq:2}
\nonumber
 \frac{\sqrt{\eta}}{f(\eta)}\left(\frac{M}{M_{Pl}}\right)^2\left(\frac{M}{T_0}\right)^{f(\eta)} &<& 10^{-5} \\
\stackrel{\eta \ll 1}{\Longrightarrow}~~~~~~~~~
\frac{1}{\sqrt{\eta}}\left(\frac{M}{T_0}\right)^{\eta} &<& 10^{-5}\left(\frac{M_{Pl}}{M}\right)^2\,,
\end{eqnarray}
or $\eta \; \lsim\; 0.01-2.5$ for $M \approx 10^{15}\;{\rm GeV}-1$
TeV. 

The general expectation in spontaneously broken
supergravity is that $\eta$ should be of
order unity. This is indeed what has 
been found so far in attempts to
embed inflation within string theory~\cite{mald}.
Thus, the above conditions on $\eta$ amount to a non-trivial
tuning that is required for phenomenological viable scenarios of 
locked inflation with or without a subsequent phase of saddle inflation.

Case ii) does allow for $\eta\sim {\cal O}(1)$ if $M$ is of order TeV scale. This is encouraging for attempts to embed inflation in string theory. 
It is important to keep in mind, however, that this assumes that density perturbations with the correct amplitude can be generated via the DGZ mechanism, even for such a low reheating temperature.
For larger values of $M$, the model is forced towards $\eta\ll 1$, corresponding to the slow-roll inflationary regime.  

\bigskip

We conclude with a comment on observational consequences of locked inflation. Here i) is the interesting case, since for case ii) the secondary ``Long Saddle Inflation'' phase erases all observational characteristics of the locked period. Locked inflation without subsequent saddle inflation either solves the standard
cosmological problems in a single step with a low
reheating scale or in a series of steps 
with a higher inflation scale, 
when $N_{locked}$ (Eq.~(\ref{Nlocked})) for each phase of inflation is
too small on its own. 
In either case one still has to satisfy the
constraint on $\eta$ in Eq.~(\ref{eq:1}) 
{\em at each step.}
In multistage locked inflation
this constraint 
does become weaker as $M$ increases, as can be seen from Fig.~\ref{cons2}. 
But the cost of this is that the constraint must be satisfied at the
end of each of the multiple phases of locked inflation.
Instead of needing to solve a stringent constraint once as is the case with
locked inflation at low mass scale, one must solve a weaker constraint
several times in order to construct a viable model.    

If there are several phases of locked inflation, this could have
immediate observational consequences. In DK's proposal, perturbations
produced during locked inflation arise via the previously mentioned DGZ mechanism, 
whereby the amplitude of the perturbations depends on the coupling of
the inflaton field to another field.  In general, this coupling and
the resulting spectrum of perturbations will have a different
amplitude during each phase of locked inflation. Consequently, the
resulting perturbation spectrum could contain discontinuities
corresponding to the different phases of locked inflation.  While
there is no guarantee that one of these discontinuities will appear in
the portion of the spectrum probed by observations, if
$M$ is large enough the amount of inflation produced during each phase
is small enough to make this unavoidable.  This argument is similar to
that considered in~\cite{Adams:2001vc} for a single field model where
the potential contains a number of steps -- putting one  ``feature''
in the perturbation spectrum requires tuning, whereas adding many
features makes it likely that at least
one will fall in the range of $k$ accessible to cosmological
measurements. This topic deserves further study,
particularly if the evidence for a running spectral index seen
in the first year data of WMAP survives.  

\section*{Acknowledgements}
We thank C.~Burgess, G.~Dvali, A.~Hamilton, S.~Kachru, A.~Linde and E.~Weinberg for useful
discussions. This work is supported in part by the DOE grants
DE-FG02-92ER-40704 (RE) and DE-FG02-92ER-40699 (KS), the CU Academic
Quality Fund and the Ohrstrom Foundation (JK).


\begin{thebibliography}{99}
\bibitem{note} This is, strictly speaking, only in true in broken
  supergravities where  
the cosmological constant is the supersymmetry breaking scale; $\Lambda
  \sim M_{susy}^4$.
\bibitem{compare} 
This can easily be mapped to DK's notation:  we can set $\alpha M_{\star}^4/2M^4 \equiv 1$ without loss of generality and have defined $\eta \equiv 4 M_{Pl}^2/M_{\star}^2$.
\end{thebibliography}
\end{document}